\documentclass[twocolumn,showpacs,pra, aps, 10pt]{revtex4-1}   
\usepackage[utf8x]{inputenc}

\usepackage{amssymb}
\usepackage{amsmath}
\usepackage{graphicx}
\usepackage{epstopdf}
\usepackage{trfsigns}
\usepackage{dsfont}
\usepackage{bm}
\usepackage{siunitx}
\usepackage{color}
\usepackage{soul}
\usepackage{cancel}

\usepackage{hyperref}

\def\fakebold#1{\relax\ifvmode\leavevmode\fi\ifmmode\setbox0=\hbox{$#1$}\else\setbox0=\hbox{#1}\fi\kern-.02em\copy0 \kern-\wd0\kern .04em\copy0 \kern-\wd0\kern-.0125em\raise.02em\box0}

\newcommand{\dif}{\,\mbox{d}}
\newcommand{\diff}{\,\mbox{\footnotesize d}}
\newcommand{\ee}{\,\mbox{e}}
\newcommand{\Tr}{\,\mbox{Tr}}

\begin{document}

\title{Canonical c-field approach to interacting Bose gases: stochastic interference of matter waves}
\author{Holger Hauptmann$^{\dagger }$}
\author{Walter T. Strunz$^{\dagger }$}
\affiliation{$^{\dagger }$Institut f\"{u}r Theoretische Physik, Technische Universit\"{a}t Dresden, 01062 Dresden, Germany}

\begin{abstract}

We present a stochastic matter field equation for an interacting many-body Bose system in equilibrium at ultracold finite temperature.
Moreover, the proposed equation can be used for non-equilibrium dynamics on phenomenological grounds.
This stochastic differential equation is based on a field phase space representation reflecting the underlying canonical density operator (fixed particle number N).
Remarkably, it allows for an efficient numerical implementation.
We apply our canonical c-field method to interference experiments with quasi-one dimensional Bose gases.
Crucial features of these interference patterns are reproduced very well and also statistical properties in terms of distributions, e.g. for the contrasts, agree well with experimental results.

\end{abstract}

\maketitle

\section{Introduction}
Understanding properties and dynamics of interacting quantum many-body systems at ultracold finite temperature is a central issue of current research.
Quantum phenomena like superfluidity are affected by mutual interactions of the constituents, thus interactions are key features for emergent phenomena \cite{leggett2006quantum, stringari, proukakis2013, pethick2002}.
While experiments with ultracold atomic gases are long established and continue to contribute to our understanding of interacting quantum many-body systems, in recent years Bose-Einstein condensates of polaritons also became an excellent playground to examine macroscopic many-body quantum effects \cite{Lerario2017, szyman2017, Juggins2018, Caputo2017, sun2017}.

A Bose-Einstein condensate consisting of weakly interacting particles ($n a^3\ll 1$ with density $n$ and s-wave scattering length $a$) at zero temperature is often treated in a mean-field approximation with the Gross-Pitaevskii equation (GPE) \cite{gross1961, pitaevskii1961}.
This nonlinear Schrödinger equation is an equation of motion for a complex classical field to describe the macroscopically occupied condensate wave function of a trapped degenerate Bose gas with two-body interaction in s-wave scattering approximation parameterized by $a$.

Beyond mean-field, Bogoliubov theory studies excitations of the condensate through expansion of the bosonic field operator around the condensate wave function \cite{bogol, parkins1981, fetter1996}.
The fluctuations of the field operators have to be small compared to the condensate, so this approach is restricted to very low temperatures $T\ll T_c$.
The Bogoliubov transformation leads to a quasiparticle description \cite{dodd, hutch} of the noncondensed fraction.
There is an extension to particle number conserving Bogoliubov theory where the field operators act on the $N$-particle subspace \cite{gardiner1997, billam, castin1998}.

A two-fluid simplification for a partially condensed Bose gas at finite temperature, when the higher modes are also highly occupied, is obtained from a Hartree-Fock approximation which leads to an extended Gross-Pitaevskii equation for the condensate \cite{huse}.
The noncondensed particles in the higher modes are described as the thermal fraction in semi-classical approximation \cite{string1996, stringari}.

The wide area of c-field methods is a further successful approach beyond mean-field Gross-Pitaevskii theory (nice overviews can be found in \cite{proukakis2013,  gardiner2008}).
The condensate and the non-condensed particles are treated in a unified description by one classical complex \textit{stochastic} field.
These theories comprise the Stochastic Projected GPE (SPGPE), the Projected GPE (PGPE) and the truncated Wigner Projected GPE (TWPGPE).
The formalisms distinguish two different parts of the field operator separated by a certain threshold energy.
The coherent band embraces low-energy modes which are highly occupied and is characterized by one classical complex field.
The high-energy band, the incoherent part, is treated quantum-mechanically.
The SPGPE governs the stochastic field which represents an open quantum system \cite{blakie2012, gardiner2002, davis2001, gardiner2003}.
The high-energy part acts as an environment which is characterized by stochastic noise.
As a simplification, neglecting the coupling between coherent and incoherent band yields the PGPE \cite{davis2005, davis2002} where the low-lying modes are considered as a closed system.

The TWPGPE for very low temperatures $T\ll T_c$ is a method similar to the truncated Wigner approximation, but using a projection operator onto the coherent band.
Stochastic sampling of the Wigner distribution yields a set of single realizations of c-fields whose time evolutions are governed by a PGPE.

Another c-field approach is the stochastic Gross-Pitaevskii equation (SGPE) derived with non-equilibrium Keldysh theory \cite{stoof1999, proukakis2013}.
Condensed and noncondensed particles are treated with one stochastic complex field.
A separation of a condensed phase is possible within the Penrose-Onsager scheme \cite{penrose}.
 
An issue of many stochastic c-field methods is the appearance of a white noise field leading to unphysical high energy contributions which need to be cut off.
In this article we present a further c-field method with two distinctive features: First, the problem of the energy cutoff is avoided in a self-contained way by colored noise in the stochastic equation.
Second, we determine the thermal state of an interacting Bose gas with fixed particle number $N$ (canonical ensemble).
In contrast to our earlier attempts \cite{sigi2009, sigi2010, sigi2013}, the stochastic matter field equation we present in this work is far more feasible for numerical implementation.

This paper has the following structure.
In the beginning, we recapitulate very briefly the damped quantum oscillator in section \ref{subsec:dampedquantumosci}.
The Glauber-Sudarshan-P representation for the density operator of an {\it{ideal}} Bose gas at ultracold finite temperature is governed by a Fokker-Planck equation which can be derived for the grand canonical ensemble from a Lindblad equation.
The corresponding It\^o equation represents an alternative point of view, where correlation functions of the stochastic matter fields are expectation values of the normal ordered field creation and annihilation operators, presented in section \ref{subsec:idealbg}.
For the canonical ensemble (particle number and temperature fixed), however, a Lindblad equation is not known. 
Nevertheless, projecting the grand canonical density operator onto the N-particle subspace, in section \ref{sec:caan} we present an exact new stochastic differential equation for the canonical ensemble of noninteracting bosons at finite temperature.
We derive an analogous canonical probability function $W_N$ and therefore an It\^o equation similar to the grand canonical case.

Interaction in the s-wave approximation takes places locally in position space therefore we switch to position representation in section \ref{subsec:posrep}.
In section \ref{sec:interactio} we phenomenologically insert self-interaction in mean-field approximation similar to the approach of the Gross-Pitaevskii equation.
Earlier work on a closely related canonical c-field method can be found in \cite{sigi2009, sigi2010, sigi2013}.
The crucial improvement of the new stochastic equation presented here is that the numerical implementation is straightforward.
We apply our new stochastic matter field equation to interference experiments with quasi-one dimensional Bose gases \cite{schmied2007, schmied2008}.
With our stochastic matter field equation we generate interference patterns in section \ref{sec:muster}, and in section \ref{sec:statis} we model contrast statistics to get very similar results to experimental outcomes \cite{schmied2007, schmied2008}.
We close the article with a discussion of the results and an outlook.

\section{The grand canonical ideal gas}
\subsection{Dynamics of a single mode Bose gas}
\label{subsec:dampedquantumosci}
Non-equilibrium dynamics of an ideal single mode Bose gas corresponds to $\hat{H}=E \hat{a}^\dagger \hat{a}$, that can be described with the help of a (Lindblad) master equation
\begin{align}
\label{lindbladmaster}
\begin{split}
 \frac{\dif \hat{\varrho}}{\dif t} = -\frac{\mathrm{i}}{\hbar} \left[ \hat{H} - \mu \hat{N}, \hat{\varrho}\right]  &+ \frac{\gamma}{2} \bar{n} \left( \left[ \hat{a}^\dagger \hat{\varrho}, \hat{a}  \right] + \left[ \hat{a}^\dagger , \hat{\varrho} \hat{a}  \right] \right) \\
  &+\frac{\gamma}{2} (\bar{n}+1) \left( \left[\hat{a} \hat{\varrho}, \hat{a}^\dagger \right] + \left[ \hat{a}, \hat{\varrho} \hat{a}^\dagger \right] \right)
\end{split}
\end{align}
with a (phenomenological) damping rate $\gamma$.
The first term on the right hand side is the von-Neumann contribution which handles the unitary dynamics.
We include a term involving the chemical potential $\mu$ and the number operator $\hat{N}=\hat{a}^\dagger \hat{a}$ to indicate the relation to the grand canonical ensemble.
The influence of the particle and thermal bath is modeled by the additional terms which describe transitions between gas and reservoir.
In the long time limit, as $t \rightarrow \infty$, this density operator reaches the equilibrium state $\hat{\varrho}(t \rightarrow \infty)=\frac{\ee^{-\beta (\hat{H}-\mu \hat{N})}}{Z}$ where the partition function $Z=\Tr(\ee^{-\beta (\hat{H}-\mu \hat{N})})$ ensures normalization.
The chemical potential adjusts the mean particle number $\bar{n}=\frac{1}{\ee^{\beta (E-\mu)}-1}$ at equilibrium.
The density operator can be depicted with the Glauber-Sudarshan-P representation in the form $\hat{\varrho}=\int \frac{\dif^2 z}{\pi} P(z,z^*) |z \rangle \langle z| $ with coherent states satisfying the eigenvalue equation $\hat{a} |z\rangle = z | z \rangle$. 
The master equation \eqref{lindbladmaster} becomes a Fokker-Planck equation
\begin{align}
\label{fokker}
\begin{split}
& \dot{P} = \left( \left(\frac{\gamma}{2}+\mathrm{i} \frac{E-\mu}{\hbar} \right) \frac{\partial }{\partial z} z +c.c.+\gamma \bar{n}  \frac{\partial^2 }{\partial z^* \partial z} \right) P \ \text{,}
\end{split}
 \end{align}
which is a differential equation for the P-function.
At equilibrium, this P-function for the thermal state $\hat{\varrho}(t \rightarrow \infty)$ (that means $\dot{P}=0$) is the Gaussian
\begin{align}
\label{Pfunct}
 P(z, z^*)=\frac{1}{\bar{n}} \ee^{-\frac{|z|^2}{\bar{n}}}=(\ee^{\beta(E-\mu)}-1)\ee^{-(\ee^{\beta(E-\mu)}-1) |z|^2} \ \text{.}
\end{align}
Generally, for a Fokker-Planck equation a corresponding Langevin equation can be found, so the It\^o stochastic differential equation
\begin{align}
\label{ito}
 \dif z= -\left( \mathrm{i} \frac{(E-\mu)}{\hbar} + \frac{\gamma}{2} \right) z \dif t +\sqrt{\gamma \bar{n} } \dif \xi
\end{align}
is the `microscopic' counterpart of equation \eqref{fokker}.
The Wiener process has the properties $\langle \dif \xi(t) \rangle=\langle \dif \xi(t) \dif \xi(t) \rangle=0$ and $\langle \dif \xi(t) \dif \xi^*(t') \rangle=\delta(t-t') \dif t$.
In the limit $\gamma=0$ this equation simplifies to $\dot{z}=- \mathrm{i} \frac{(E-\mu)}{\hbar} z$ and the coherent state label $z(t)=z(0) \ee^{-\mathrm{i} \frac{(E-\mu) t}{\hbar}}$ rotates in the imaginary plane due to their eigenenergy which is consistent with the time evolution of coherent states in an oscillator \cite{schleich2001}.
While the frequency $\frac{E-\mu}{\hbar}$ determines the time scale of the unitary rotation in phase space, the rate $\gamma$ sets the time scale of amplitude damping.
Equation \eqref{ito} is an Ornstein-Uhlenbeck process for the coherent state label $z$.

Normal ordered expectation values of creation and annihilation operators $\langle \hat{a}^\dagger \hat{a}^\dagger \ldots \hat{a} \hat{a} \rangle$ can either be interpreted as moments of the P-function or ensemble means of paths $z(t)$ of the It\^o equation
\begin{align}
\label{expectationvalue}
\begin{split}
  \Tr( \hat{a}^\dagger \hat{a}^\dagger \ldots \hat{a} \hat{a} \ \hat{\varrho} ) &= \int \frac{\dif^2 z}{\pi} \ z^* z^* \ldots z z \ P(z, z^*) \\
  &= \langle \hspace{-0.2cm} \langle z^*(t) z^*(t) \ldots z(t) z(t) \rangle \hspace{-0.2cm} \rangle
\end{split}
\end{align}
where the brackets $\langle \hspace{-0.2cm} \langle \, \ldots \, \rangle \hspace{-0.2cm} \rangle$ denote ensemble means of independent solutions of the It\^o equation \eqref{ito}.
At equilibrium (long time limit $\gamma t \gg 1$) this ensemble mean is identical to the time average over one single path $\langle \hspace{-0.2cm} \langle \, \ldots \, \rangle \hspace{-0.2cm} \rangle=\lim\limits_{T\rightarrow \infty}{\frac{1}{T} \int_0^T \dif t (\ldots)}$.

\subsection{Ideal gas in the grand canonical ensemble}
\label{subsec:idealbg}
Here we summarize the procedure how to arrive at an It\^o stochastic differential equation for Bose gases and how to calculate correlation functions of arbitrary order.
Temporarily we digress to an ideal gas (without self interaction) with damping in the grand canonical ensemble in order to understand the approach for the interacting gas in the canonical ensemble.
In second quantization the Hamiltonian for an ideal Bose gas in energy representation can be written as
\begin{align}
\label{hamiltonian}
 \hat{H}=\sum_i \limits E_i \hat{a}_i^\dagger \hat{a}_i =: \sum_i \limits \hat{H}_i
\end{align}
which is the sum of a combination of creation operators $\hat{a}_i^\dagger$ and annihilation operators $\hat{a}_i$ over all trap modes $i$.
These operators obey the bosonic commutation relations $[\hat{a}_j, \hat{a}_k^\dagger]=\delta_{jk}$ and the eigenvalue equation $\hat{a}_i |z_i \rangle= z_i |z_i \rangle$ for coherent states $| z_i \rangle$ is fulfilled.
At equilibrium the density operator for the grand canonical ensemble is
\begin{align}
\label{rhocan}
 \hat{\varrho} =\frac{1}{Z} \ee^{-\beta (\hat{H}-\mu \hat{N}) }
\end{align}
with the Hamiltonian given in equation \eqref{hamiltonian} and the partition function $Z$ ensures normalization of the operator $\operatorname{Tr}(\hat{\varrho})=1$.
For the grand canonical ensemble we consider $\hat{H}-\mu \hat{N}$ in the exponential, with the chemical potential $\mu$ adjusting the mean total particle number and the number operator being $\hat{N}=\sum_i \hat{a}_i^\dagger \hat{a}_i $.
The structure of the Hamiltonian \eqref{hamiltonian} ensures that the density operator factorizes into the product 
\begin{align}
\label{totalrho}
 \hat{\varrho} =\prod_i \limits \hat{\varrho}_i
\end{align}
with density operators
\begin{align}
\label{denop}
 \hat{\varrho}_i = \frac{1}{Z_i}  \ee^{-\beta (E_i-\mu) \hat{a}^\dagger_i \hat{a}_i }
\end{align}
for each mode $i$ which are also normalized $\operatorname{Tr}(\hat{\varrho}_i)=1$.
That density operator for mode $i$ conforms with the equilibrium state of the damped quantum oscillator from the previous subsection, and so expressions of P-functions, Fokker-Planck equations and corresponding It\^o equations can be adopted (equations see Appendix \ref{sec:anhang0}).
Equivalently, representations of the operators $\hat{\varrho}_i$ in P-functions \cite{gardiner1993} are
\begin{align}
\label{definitio}
 \hat{\varrho}_i=\int \frac{\dif^2 z_i}{\pi} P_i(z_i, z_i^*) |z_i \rangle \langle z_i |
\end{align}
and the total P-function of the total density operator $\hat{\varrho}$, equation \eqref{totalrho}, is $P(\{z\}):= P(z_1,z_1^*,z_2,z_2^*,z_3,z_3^*, \ldots)=\prod_i P_i(z_i,z_i^*)$.
Finally the equivalence of Fokker-Planck equations with stochastic differential equations ensures It\^o equations for $z_i$ which read
\begin{align}
\label{gcstoch}
 \dif z_i= -\left( \mathrm{i} \frac{E_i-\mu}{\hbar} + \frac{\gamma_i}{2} \right) z_i \dif t +\sqrt{\gamma_i n_i } \dif \xi_i
\end{align}
with independent Wiener processes $\xi_i(t)$.
These stochastic c-number differential equations for the grand canonical ensemble correspond to an ensemble of Lindblad equations.
At equilibrium (functions $P_i$ are time independent) the analogue of equation \eqref{Pfunct} is accessible with inversion of equation \eqref{definitio} 
and reads $P_i(z_i,z_i^*) = (\ee^{\beta (E_i-\mu) }-1) \ee^{-|z_i|^2 ( \ee^{\beta (E_i-\mu)}-1)}$  \cite{mandel1995}. 
Finally, the total P-function from equation \eqref{Pfunct} reads
\begin{align}
\label{Pfull}
  P(\{z\}) = \left(\prod_j \limits (\ee^{\beta (E_j-\mu) }-1) \right) \ee^{-\sum_k \limits |z_k|^2 ( \ee^{\beta (E_k-\mu)}-1)}
\end{align}
and is a Gaussian distribution of the coherent state labels $z_i$ which is characteristic for an Ornstein-Uhlenbeck process.
This expression is also a solution of the Fokker-Planck equation \eqref{fpegc} in the stationary case with $\frac{\partial P_i}{\partial t}=0$.
It can be seen as a probability density function for $z_i$ and $z_i^*$, and fulfills $ \int \frac{\dif^2 z_i}{\pi} P_i(z_i,z_i^*)=1$ as a normalization property.
For later use we introduce 
\begin{align}
\label{pfunctionunnorm}
 \tilde{P}_i(z_i,z_i^*)= \ee^{\beta (E_i-\mu)} \ee^{-|z_i|^2(e^{\beta (E_i-\mu)}-1)}
\end{align}
which is the P-function for the representation of the unnormalized density operator
\begin{align}
\label{Pfunction}
 \ee^{-\beta (E_i-\mu) \hat{a}_i^\dagger \hat{a}_i} = \int \frac{\dif ^2 z_i}{\pi} \tilde{P}_i(z_i,z_i^*) | z_i \rangle \langle z_i |  \ \text{.}
\end{align}

In the following we calculate correlation functions.
As an example we begin with the first order correlation function
\begin{align}
\begin{split}
\label{firstorder}
 \langle \hat{a}^\dagger_i \hat{a}_i\rangle_\infty = \int \frac{\dif^2 z_i}{\pi} |z_i|^2 P_i(z_i,z_i^*) = \frac{1}{e^{\beta (E_i-\mu)}-1}
\end{split}
\end{align}
where we recover the well known occupation number $ n_i = \langle \hat{a}^\dagger_i \hat{a}_i\rangle_\infty=\operatorname{Tr}( \hat{a}^\dagger_i \hat{a}_i \hat{\varrho})$.
The index of the brackets $\langle \ldots \rangle_\infty$ indicates expectation values in the grand canonical ensemble.

In general it is possible to express expectation values of arbitrary order of normal-ordered products 
\begin{align}
\label{arbitorder}
\begin{split}
 \langle \hat{a}^\dagger_j \hat{a}^\dagger_k \ldots \hat{a}_l \hat{a}_m \rangle_\infty &= \int \dif^2 \{z\} \ z_j^* z_k^* \ldots z_l z_m P(\{z\}) 
\end{split}
\end{align}
as an integral over all coherent state labels.
The integration is a multi integral $\int \dif^2 \{ z \}:=\int \frac{\diff^2 z_k}{\pi} \int \frac{\diff^2 z_l}{\pi} \ldots $.
Here we emphasize that these expectation values are the corresponding moments of the probability density function $P(\{z\})$.
For not normal-ordered products the bosonic commutation relations for $\hat{a}_i$ and $\hat{a}_j^\dagger$ are available to get a sum of normal-ordered products. 
If the number of creation and annihilation operators are unequal the expectation values are $0$ because the probability density functions $P_i$ are Gaussians centered at $z_i=0$.
It is a direct consequence of equation \eqref{arbitorder} that correlation functions are mean values 
\begin{align}
\langle \hat{a}^\dagger_j \hat{a}^\dagger_k \ldots \hat{a}_l \hat{a}_m \rangle_\infty = \langle \hspace{-0.2cm} \langle \, z_j^*(t) z_k^*(t) \ldots  z_l(t) z_m(t) \, \rangle \hspace{-0.2cm} \rangle_\infty
\end{align}
of stochastic paths of equations \eqref{gcstoch}, where the brackets $\langle \hspace{-0.2cm} \langle \, \ldots \, \rangle \hspace{-0.2cm} \rangle_\infty$ denote ensemble means (or time averages in the stationary case) of independent solutions of the grand canonical stochastic differential equations \eqref{gcstoch}.
For example, in the long time limit ($\gamma_i t \gg 1$) occupation numbers $\langle \hat{a}^\dagger_j \hat{a}_j \rangle_\infty = \langle \hspace{-0.2cm} \langle \, z_j^*(t) z_j(t) \, \rangle \hspace{-0.2cm} \rangle_\infty$ can be calculated with time averages of solutions of the It\^o equation.

\section{The canonical ideal gas}
\label{sec:caan}

In this section we derive a stochastic matter field equation for the canonical ensemble consisting of $N$ bosonic particles as an analogue of the grand canonical equations \eqref{gcstoch}.
For the canonical ensemble we do not have a Lindblad equation as a starting point.
So we calculate expectation values for the canonical ensemble and find a probability density function $W_N(\{z\})$ as the canonical counterpart to the grand canonical density function $P(\{z\})$.
 
\subsection{Equilibrium distribution and expectation values}
The grand canonical density operator \eqref{denop} has to be constrained to $N$ particles so we multiply with a projection operator to get the canonical density operator
\begin{align}
\label{candenop}
 \hat{\varrho}_N = \frac{1}{Z_N} \ee^{-\beta \hat{H\,}} \hat{\Pi}_N \ \text{.}
\end{align}
Again the canonical partition function $Z_N$ ensures normalization $\operatorname{Tr}(\hat{\varrho}_N)=1$.
The operator $\hat{\Pi}_N=\sum_{n=N} | \{ n \} \rangle \langle \{ n \} |$ projects the exponential onto the $N$-particle subspace, where $|\{n\} \rangle=| n_0 n_1 n_2 \ldots n_l \ldots \rangle$ with $\sum_k n_k=N$ is an $N$-particle number state.
The Hamiltonian is again given in equation \eqref{hamiltonian}.
We calculate the canonical partition function
\begin{align}
\label{canzusum}
\begin{split}
 Z_N  & =\operatorname{Tr} \left(\ee^{-\beta \sum_i \limits E_i \hat{a}^\dagger_i \hat{a}_i} \hat{\Pi}_N \right) \\
  & = \prod_j \ee^{\beta E_j} \int \dif^2 \{ z \} \frac{1}{N!}  \left( \sum_l \limits |z_l|^2 \right)^{N} \ee^{-\sum_l \limits |z_l|^2 \ee^{\beta E_l}} \\
  & =:\prod_j \ee^{\beta E_j} \int \dif^2 \{ z \} \ W_N(\{ z \})
 \end{split}
\end{align}
where we introduce the weight function
\begin{align}
 \label{weighting}
 W_N(\{ z \} )= \frac{1}{N!}\left( \sum_l \limits |z_l|^2 \right)^{N} \ee^{-\sum_l \limits |z_l|^2 \ee^{\beta E_l}} 
\end{align}
as a main result.
Details for the calculation can be found in Appendix \ref{sec:anhangdetails}.
This weight function depends on all coherent state labels, which we mark with $W_N(\{ z \})$.
The correlation functions in the canonical ensemble are expectation values using the canonical density operator, equation \eqref{candenop}.
As an example, the first order correlation function leads to a multi integral expression
\begin{align}
\label{firstorder3}
 \langle \hat{a}^\dagger_j \hat{a}_j\rangle_N = N \  \frac{\int \dif^2 \{ z \} \ \Big( \frac{|z_j|^2}{\sum_l \limits  |z_l|^2} \Big) \ W_{N} (\{ z \} )}{\int \dif^2 \{ z \} \ W_N(\{ z \})} \ \text{.}
\end{align}
Expectation values of arbitrary normal-ordered products can be expressed as
\begin{align}
\label{generalorder3}
\begin{split}
 \langle \underbrace{ \hat{a}^\dagger_j \hat{a}^\dagger_k \ldots \hat{a}_l  \hat{a}_m }_{\substack{2M}}&\rangle_N =  \frac{N!}{(N-M)!}\frac{1}{\int \dif^2 \{ z \} \ W_N(\{ z \})} \cdot \\
 & \int \dif^2 \{ z \} \ \Bigg( \frac{z_j^* z_k^* \ldots z_l z_m}{ \big(\sum_p \limits |z_p|^2\big)^M} \Bigg) \ W_{N} (\{ z \} )
\end{split}
\end{align}
with $N\geq M$ (equation \eqref{generalorder2} in Appendix \ref{sec:anhangdetails}).
The brackets $\langle \ldots \rangle_N$ indicate expectation values in the canonical ensemble for $N$ particles.

\subsection{New stochastic matter field equation}

So far we expressed canonical correlation functions as integrals over coherent state labels.
Similarly to the P-functions in the grand canonical ensemble we can interpret the weight function $W_N$ as a stationary probability density function (which is not normalized to $1$). 
Moreover, the correlation functions are moments of the weight function.
The integrals in equations \eqref{firstorder3} and \eqref{generalorder3} can be seen as Monte-Carlo integrals, with the probability density function $W_N$.
This stationary probability density function $W_N$ then should obey a stationary Fokker-Planck equation
\begin{align}
\label{fpewn}
\begin{split}
 0= \sum_i \limits\left( \frac{\partial }{\partial z_i} A_i z_i + \frac{\partial }{\partial z_i^*} A_i^* z_i^* +B_{i}  \frac{\partial^2 }{\partial z_i^* \partial z_i} \right) W_N(\{z \})
\end{split}
\end{align}
which has the same form as the grand canonical counterpart, equation \eqref{fokker} or respectively equation \eqref{fpegc}.
For now, the coefficients for the drift $A_i$ and diffusion $B_i$ are unknown.
Different from equation \eqref{fokker} or equation \eqref{fpegc}, where these coefficients are known and we look for a solution $P(z)$ or $P_i(z_i, z_i^*)$, here in equation \eqref{fpewn} we search for coefficients $A_i$ and $B_i$ belonging to an identified $W_N$, equation \eqref{weighting}.
One can prove (see also appendix) that one possible choice of the drift and diffusion is
\begin{align}
\label{parameter}
\begin{split}
 A_i &=\mathrm{i} \frac{E_i}{\hbar} + \frac{\Lambda_i}{2} - \frac{\Lambda_i}{2} \frac{N \ee^{-\beta E_i}}{\sum_l\limits |z_l|^2} \\
 B_i &=\Lambda_i \ee^{- \beta E_i}
\end{split}
\end{align}
such that $W_N$ from equation \eqref{weighting} is a solution of equation \eqref{fpewn}.
This is similar to the spirit of \cite{sigi2009, sigi2010, sigi2013}, where a square root of the Hamiltonian in the fluctuations required sophisticated numerical implementation.
In equation \eqref{parameter} we exploit the freedom to find other drift and diffusion terms which obey equation \eqref{fpewn} and lead to much improved numerical performance.
As in the grand canonical case, the damping parameters $\Lambda_i$ determine the time scale of the damping dynamics.
The corresponding coupled set of It\^o stochastic differential equations for the coherent state labels
\begin{align}
\label{canstoch}
 \dif z_i=-A_i z_i \dif t+\sqrt{B_i} \dif \xi_i
\end{align}
is an important result and the canonical counterpart of the grand canonical equations \eqref{gcstoch}.
With these stochastic equations it is possible to get a set of independent samples $\{ z_l \}$ in the long time limit which recover the probability density function $W_N$.
Then the correlation functions, equations \eqref{firstorder3} and \eqref{generalorder3}, are mean values (notation $\langle \hspace{-0.1cm} \langle \ldots \rangle \hspace{-0.1cm} \rangle_N$)
\begin{align}
\label{firstorder3a}
\begin{split}
\langle \hat{a}^\dagger_j \hat{a}_j\rangle_N &= N \Bigg\langle \hspace{-0.2cm} \Bigg\langle  \frac{|z_j|^2}{\sum_l \limits |z_l|^2}  \Bigg\rangle \hspace{-0.2cm} \Bigg\rangle_N 
\end{split}
\end{align}
and for $N\geq M$
\begin{align}
\label{generalorder3a}
\begin{split}
\langle \underbrace{ \hat{a}^\dagger_j \hat{a}^\dagger_k \ldots \hat{a}_l  \hat{a}_m }_{\substack{2M}}\rangle_N &= \frac{N!}{(N-M)!} \Bigg\langle \hspace{-0.2cm} \Bigg\langle   \frac{z_j^* z_k^* \ldots z_l z_m}{ \big(\sum_p \limits |z_p|^2\big)^M}  \Bigg\rangle \hspace{-0.2cm} \Bigg\rangle_N  
\end{split}
\end{align}
of independent relaxed ($\Lambda_i t \gg 1$) solutions of the canonical stochastic differential equations \eqref{canstoch}.

\subsection{Position representation}
\label{subsec:posrep}
Up to now the derivations were done in energy representation, which is the natural choice for noninteracting particles.
We would like to include self-interaction in s-wave scattering approximation.
This contact interaction takes place in position space, which is the motivation to switch to position representation or basis independent bra-ket notation.
The Hamiltonian \eqref{hamiltonian} in second quantization in position representation
\begin{align*}
 \hat{H\,}= \int \dif x \ \hat{\Psi}^\dagger(x) \left(-\frac{\hbar^2}{2 m} \bigtriangleup + V(x) \right)\hat{\Psi}(x) 
\end{align*}
can be expressed with bosonic field operators.
These fulfill the commutation relations $[\hat{\Psi}(x), \hat{\Psi}^\dagger(x') ]=\delta(x-x')$, $[\hat{\Psi}(x), \hat{\Psi}(x') ]=[\hat{\Psi}^\dagger(x), \hat{\Psi}^\dagger(x') ]=0$.
We can expand the grand canonical density operator using projectors of the coherent field states $| \psi \rangle$, with $\hat{\Psi}(x) | \psi \rangle= \psi(x)| \psi \rangle$, as
\begin{align}
\label{fullfunctional}
  \frac{1}{Z} \ee^{-\beta \hat{H \,}} = \frac{\int {\cal{D}} [\psi,\psi^*] \  P[\psi, \psi^*] \ | \psi \rangle \langle \psi |}{\int {\cal{D}} [\psi,\psi^*] \  P[\psi, \psi^*]} \ \text{,}
\end{align}
where the integral becomes a functional integral $\int {\cal{D}} [\psi,\psi^*] \ldots$ over $\psi$ and $\psi^*$.
Here the P-functional $P[\psi, \psi^*]$ needs not be normalized but normalization of the whole expression \eqref{fullfunctional} takes place because we divide by the zeroth moment.
Neglecting normalization, the P-function in equation \eqref{Pfull}, becomes the functional
\begin{align}
\label{Pfunctional}
\begin{split}
 P[\psi,\psi^*]&= \ee^{-\langle \psi| \ee^{\beta (H-\mu)}-1 | \psi \rangle} \\
                      &=\ee^{-\int \diff x \int \diff x' \ \psi^*(x) (\ee^{\beta (H(x)-\mu)}-1) \psi(x')}
\end{split}
\end{align}
in bra-ket notation and position representation with  $H(x)= -\frac{\hbar^2}{2 m} \bigtriangleup + V(x)$.
Analogously, the probability density function $W_N$ for the canonical ensemble, equation \eqref{weighting}, transforms to the functional
\begin{align}
\label{Wfunctional}
 \begin{split}
  & W_N[\psi, \psi^* ] =\frac{1}{N!} (\langle \psi | \psi \rangle )^N \ee^{-\langle  \psi | \ee^{\beta H} | \psi \rangle} \\
  &=\frac{1}{N!} \left( \int |\psi(x)|^2 \dif x \right)^N \ee^{-\int \diff x \int \diff x' \ \psi^*(x) \ee^{\beta H(x)} \psi(x')}
 \end{split}
\end{align}
of $\psi$ and $\psi^*$.
When we respect the gauge $H-\mu \leftrightarrow H$, the unnormalized probability density functionals of the grand canonical ensemble $P[\psi, \psi^*]$ and of the canonical ensemble $W_N[\psi, \psi^*]$ are connected to each other
\begin{align*}
 \begin{split}
   W_N [\psi, \psi^* ] & = \frac{1}{N!} \left( \langle \psi | \psi \rangle \right)^N e^{-\langle \psi | \psi \rangle} P[\psi, \psi^*] \\
   \sum_{N=0 }^\infty \limits W_N [\psi, \psi^* ] & = P[\psi, \psi^*] \text{.}
\end{split}
\end{align*}
As an example, we show expressions for the densities for the grand canonical ensemble and canonical ensemble in position representation in Appendix \ref{sec:anhangdensities}.

Both unnormalized probability density functionals $Q=(P[\psi, \psi^* ] , W_N[\psi, \psi^* ])$ obey a stationary functional Fokker-Planck equation
\begin{align}
\label{fpefunctional}
\begin{split}
 0 =\frac{\partial Q}{\partial t} = \int \hspace{-0.2cm} \dif x & \frac{\delta ( A \psi  Q) }{\delta \psi(x)} + c.c.+ \int \hspace{-0.2cm} \dif x \int \hspace{-0.2cm} \dif x'  \frac{\delta^2 (B Q)}{ \delta \psi(x) \delta \psi^*(x')}
\end{split}
\end{align}
in functional notation with coefficients $A[\psi,\psi^*](x)$ and $B[\psi,\psi^*](x,x')$ \cite{zubarev1997}.
For the grand canonical ensemble these coefficients
\begin{align}
\begin{split}
 A_{\infty}[\psi,\psi^*](x)&=\left( \mathrm{i} \frac{ H(x)-\mu}{\hbar} + \frac{\gamma(x)}{2} \right)  \\
 B_{\infty}[\psi,\psi^*](x,x') &=\langle x | \gamma \big( \ee^{\beta (H-\mu)}-1\big)^{-1}|x'\rangle
\end{split}
\end{align}
are well known from equations \eqref{gcstoch}.
In appendix \ref{sec:anhang1} we give some intermediate steps to prove that $P[\psi, \psi^*]$ is a solution of the Fokker-Planck equation \eqref{fpefunctional}.
Finally we find \cite{Breuer2007} a corresponding It\^o equation
\begin{align}
\label{gcpos}
\begin{split}
  \dif \psi(x,t) =- &\left( \mathrm{i}  \frac{ H(x)-\mu}{\hbar} + \frac{\gamma}{2} \right) \psi(x,t) \dif t 
  \\&+ \sqrt{\frac{ \gamma}{\ee^{\beta (H(x)-\mu)}-1 } }  \dif \xi(x,t)
\end{split}  
\end{align}
which is the position representation of equations \eqref{gcstoch}.

As in energy representation the coefficients $A_{N}[\psi,\psi^*]$ and $B_{N}[\psi,\psi^*](x,x')$ have to be determined.
However, an expression for $W_N$, equation \eqref{Wfunctional}, was derived from expectation values and can be seen as a the solution of a Fokker-Planck equation.
Just as in energy representation, equation \eqref{parameter}, we set
\begin{align}
\label{canparameter}
\begin{split}
 &  A_N(x)= \mathrm{i} \frac{H}{\hbar} +\frac{\Lambda}{2} - \frac{\Lambda}{2} \frac{N \ee^{-\beta H}}{\langle \psi| \psi\rangle} \\
 &  B_N(x,x')=\big\langle x \big|   \Lambda  \ee^{-\beta H}  \big| x' \big\rangle \ \text{.}
\end{split}
\end{align}
In appendix \ref{sec:anhang2} we show some steps to calculate that $W_N$ solves the Fokker-Planck equation \eqref{fpefunctional} with these coefficients $A_N$ and $B_N$.

One of our main results of this paper is the It\^o stochastic differential equation
\begin{align}
 \label{SMFE}
 \begin{split}
 \dif \psi(x, t) &= \left(-\mathrm{i} \frac{H}{\hbar} -\frac{\Lambda}{2} + \frac{\Lambda}{2} \frac{N \ee^{-\beta H}}{\langle \psi| \psi\rangle}  \right) \psi(x,t) \dif t  \\
 &+\sqrt{ \Lambda}  \ee^{-\frac{1}{2}\beta H}   \dif \xi(x,t) 
 \end{split}
\end{align}
for the canonical ensemble.
This canonical stochastic matter field equation is equation \eqref{canstoch} in position representation and corresponds to equation \eqref{gcpos} for the grand canonical case.
Note, that the norm $\langle \psi | \psi \rangle$ in the drift term causes a `global' coupling of the matter fields between all points in position space.
Also note, that equation \eqref{SMFE} is different from earlier work \cite{sigi2009, sigi2010, sigi2013}, especially a square root of the Hamiltonian is avoided.
This new stochastic matter field equation can simply be treated numerically with the split-operator method (also for spatially three dimensional, spherically asymmetric Bose gases).

\section{Self interacting Bose gas}
\label{sec:interactio}
Until now we derived exact stochastic differential equations for the noninteracting Bose gas in different ensembles.
In most experiments the Bosons influence each other so it is essential to extend our description to interacting particles.
Unfortunately, expressions for $P$ and $W_N$ are not known for the interacting case.

The standard approach is the s-wave scattering approximation which treats low-energy contact interaction in position space.
The Hamiltonian in second quantization can be written as
\begin{align*}
  \hat{H\,} &= \int \dif x \ \bigg( \hat{\Psi}^\dagger(x)  \left(-\frac{\hbar^2}{2 m} \bigtriangleup + V(x) \right)\hat{\Psi}(x) \\
  & \ \ \ \ \  \ \ \ \ \ \ \ \ \ \ + \frac{g}{2} \hat{\Psi}^\dagger(x) \hat{\Psi}^\dagger(x) \hat{\Psi}(x) \hat{\Psi}(x) \bigg) 
\end{align*}
with a self interaction strength $g=\frac{4 \pi a \hbar^2}{m}$ proportional to the scattering length $a$ .
In mean-field approximation this leads to Gross-Pitaevskii theory with an effective 'Hamiltonian'
\begin{align*}
 H(x)=-\frac{\hbar^2}{2m} \bigtriangleup +V(x)+g N |\psi(x)|^2 \ \text{.}
\end{align*}

In the same way we claim to include self interaction phenomenologically to our theory by replacing the free Hamiltonian in equation \eqref{SMFE} by 
\begin{align}
\label{GPH}
 H(x)=-\frac{\hbar^2}{2m} \bigtriangleup +V(x)+gN\frac{|\psi(x)|^2}{\langle \psi | \psi \rangle}
\end{align}
as an important approximation.
Here we have to renormalize the interaction term, because stochastic solutions of equation \eqref{SMFE} are unnormalized.
Hence we always use Hamiltonian \eqref{GPH} in our stochastic matter field equation \eqref{SMFE}.

The derivation of this equation is exact without self interaction ($g=0$).
Finally, the stochastic matter field equation \eqref{SMFE} with Hamiltonian \eqref{GPH} can be seen as a stochastic Gross-Pitaevskii equation for the \textit{canonical} ensemble.
A single realization of the stochastic matter field $\psi(x)$ is unnormalized.
This lack of normalization of this state $\psi$ is not problematic.
We find that the calculation of arbitrary correlation functions ($N \geq M$)
\begin{align*}
 &\langle \underbrace{\hat{\Psi}^\dagger(x) \hat{\Psi}^\dagger(x') \ldots \hat{\Psi}(x'') \hat{\Psi}(x''')  }_{\substack{2M}}\rangle_N  \\
 &\ \ \ \ = \frac{N!}{(N-M)!} \Bigg\langle \hspace{-0.2cm} \Bigg\langle  \frac{\psi^*(x) \psi^*(x') \ldots \psi(x'')\psi(x''')}{\big( \langle \psi | \psi \rangle \big)^M}  \Bigg\rangle \hspace{-0.2cm} \Bigg\rangle_N 
\end{align*}
with sample means over many independent stochastic solutions of equation \eqref{SMFE} involves a normalizing denominator to ensure agreement with quantum expectation values.
We have access to the full many body quantum state of interacting particles because we can calculate correlation functions of arbitrary order.
That means that we can calculate arbitrary moments of the functional $W_N$ which is equivalent with knowing the probability functional itself.

\begin{figure}[htbp]
\begin{minipage}{8.5cm}
 \includegraphics[height=4.7cm, width=8.5cm]{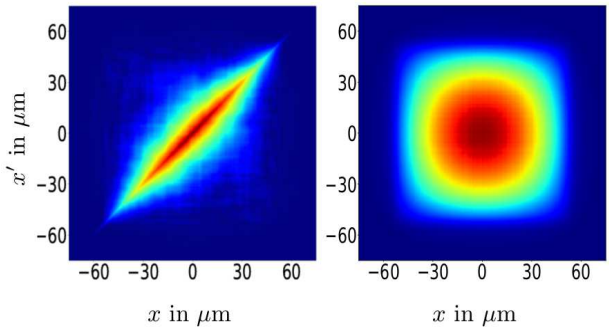} 
\end{minipage}
\caption{ Left: Absolute value of first order correlation function $|G_1(x,x')|$. Right: Absolute value of second order correlation function $|G_2(x,x')|$.
The values are $^{87}$Rb, $N=4400$, $T=\SI{31}{nK}$, radial harmonic trap frequency $\omega_\perp=2 \pi \cdot \SI{ 3000}{Hz}$, longitudinal harmonic trap frequency $\omega=2 \pi \cdot \SI{ 12}{Hz}$, averaging $\langle \hspace{-0.2cm} \langle \hspace{0.2cm} \ldots \hspace{0.2cm} \rangle \hspace{-0.2cm \rangle _N}$ over $400$ realizations.
Color scale in arbitrary units.}
\label{fig:correlationfunctions}
\end{figure}

As an example in Fig. \ref{fig:correlationfunctions} we show the absolute values of the correlation function of first order
\begin{align*}
 G_1(x,x') &= \langle \hat{\Psi}^\dagger(x) \hat{\Psi}(x') \rangle_N= N \Bigg\langle \hspace{-0.2cm} \Bigg\langle  \frac{\psi^*(x) \, \psi(x')}{\langle \psi | \psi \rangle}  \Bigg\rangle \hspace{-0.2cm} \Bigg\rangle_N 
\end{align*}
and the correlation function of second order
\begin{align*}
 G_2(x,x') &= \langle \hat{\Psi}^\dagger(x) \hat{\Psi}^\dagger(x') \hat{\Psi}(x) \hat{\Psi}(x') \rangle_N \\
 &= N(N-1) \Bigg\langle \hspace{-0.2cm} \Bigg\langle  \frac{\psi^*(x) \psi^*(x') \psi(x) \psi(x')}{(\langle \psi | \psi \rangle)^2}  \Bigg\rangle \hspace{-0.2cm} \Bigg\rangle_N 
\end{align*}
for a one dimensional self interacting Bose gas in the canonical ensemble calculated with independent solutions of our stochastic matter field equation
(for related results see \cite{gomes2006}).

The freedom $V(x)\rightarrow V(x) + \operatorname{const}$ in our stochastic matter field equation is reflected by the change of the average value of the norm of $\psi$.
Therefore, a shift in the energy results in a rescaling of the value of the norm, without affecting physical observables. 

The positive real number $\Lambda$ in equation \eqref{SMFE} is a damping rate.
We are interested in equilibrium many-body states, so $\Lambda$ is an arbitrary free parameter.
The damping rate is a measure how long we should propagate equation \eqref{SMFE} to come from non-equilibrium to an equilibrium state.
Of course, the many-body equilibrium state $\psi$ is independent of the value for $\Lambda$.

We remark that our stochastic matter field equation \eqref{SMFE} is ultimately driven by \textit{colored} noise.
The white noise $\dif \xi(x)$ is modified by an operator which includes $\psi$ itself in the Hamiltonian.
This colored noise takes care that usual high-energy fluctuations in stochastic Gross-Pitaevskii equations are suppressed and cutoff problems are prevented by construction.
Details about a similar stochastic matter field equation for the canonical ensemble, corresponding to our functional $W_N$, can be found in article \cite{sigi2013}.

As a comment we mention that the choice $\Lambda=\frac{2 k_B T}{\hbar} \gamma \ee^{\beta H}$ and the high temperature limit $\ee^{\beta H}=1+\beta H$ connects our stochastic matter field equation to a stochastic Gross-Pitaevskii equation
\begin{align}
\label{kanonischstochgrosspit}
\begin{split}
  \mathrm{i} \hbar \frac{\partial \psi}{\partial t} =& \Bigg( -\frac{\hbar^2}{2m} \bigtriangleup + V-\mathrm{i} R + gN \frac{|\psi|^2}{\langle \psi| \psi \rangle} \\
 & + \mathrm{i} \gamma k_B T \frac{(N-\langle \psi | \psi \rangle)}{\langle \psi | \psi \rangle} \Bigg) \psi(x, t) + \eta(x, t)
\end{split}
\end{align}
with $R= \gamma \left(-\frac{\hbar^2}{2m} \bigtriangleup + V(x) + gN \frac{|\psi|^2}{\langle \psi | \psi \rangle} \right)$ and white noise $\langle \eta^*(x,t) \eta(x' ,t') \rangle = 2 \hbar \gamma k_B T \delta(x-x') \delta(t-t')$.
This canonical equation looks very similar to the usual stochastic Gross-Pitaevskii equation for the grand canonical ensemble \cite{proukakis2013}.
The last term $\mathrm{i} \gamma k_B T \frac{(N-\langle \psi | \psi \rangle)}{\langle \psi | \psi \rangle}$ acts like an imaginary chemical potential, which adjusts an (unphysical) norm of the field $\psi$.
There is a wide area of grand canonical stochastic Gross-Pitaevskii equations similar to equation \eqref{kanonischstochgrosspit}, as an example applied to polariton condensates see \cite{Juggins2018, carusotto2013, matus2015}. 
In the next sections we apply our stochastic matter field equation to experiments with atomic gases.

\section{As in experiment: ensemble of stochastic patterns}
\label{sec:muster}
In experiments \cite{schmied2007, schmied2008} it is routine to interfere two independent quasi-one dimensional clouds of bosonic interacting Rubidium atoms at finite ultracold temperature.
The resulting interference patterns are composed of interference stripes which are randomly placed and staggered.
Moreover, every repetition delivers another pattern but with the same crucial features.
In contrast to interference experiments of the last centuries these latest efforts reveal interacting many-particle physics.
It is necessary to extend the description from wave functions (quantum mechanics) to quantum field operators (quantum field theory).

In this section we apply our stochastic matter field equation \eqref{SMFE} to such experiments \cite{schmied2007, schmied2008} and show the modeling of such random interference patterns in three steps.
\subsection{Preparation}
In a first step we provide two independent quasi one-dimensional stochastic matter fields $\Psi_1$ and $\Psi_2$ which are arranged parallel side by side.
For the longitudinal direction of the many-body states $\Psi_1$ and $\Psi_2$ we numerically take two one-dimensional stochastically independent solutions $\psi_1(z)$ and $\psi_2(z)$ of equation \eqref{SMFE} for finite temperature with a harmonic trap $V=\frac{1}{2} m \omega^2 z^2$.
The effective one-dimensional interaction strength $g_{1D}=2 \hbar \omega_\perp a$ is determined by the trapping frequency $\omega_\perp$ for the transverse dimensions.
We assume that the Bose gases are in harmonic oscillator ground states along the transverse narrow directions ($\omega_\perp \gg \omega$) so we take two Gaussians
\begin{align*}
\phi_{1,2} (x,y, t=0)= \left(\frac{m \omega_\perp}{\pi \hbar}\right)^{1/2} \ee^{-\frac{m \omega_\perp}{2 \hbar} \left(\left(y \pm \frac{d}{2}\right)^2+x^2\right)}
\end{align*}
at distance $d$ in $y$-direction.
The gases are in harmonic oscillator ground states along the $x$-direction, too.
Later, the absorption direction for the measurement is oriented along the $x$-axis which requires integration over the density along $x$-direction.
We use the values $d=\SI{3.5}{\micro m}$, $m=\SI{87}{u}$ and $a=\SI{5.77}{nm}$ (\cite{dalfavo}) for $^{87}$Rb, particle number $N=4400$, two different temperatures $T=\SI{31}{nK}$ and $T=\SI{60}{nK}$ and frequencies $\omega_\perp=2 \pi \cdot \SI{ 3000}{Hz}$, $\omega=2 \pi \cdot \SI{ 12}{Hz}$ which are chosen to match the experimental conditions.   
\subsection{Ballistic expansion}
\begin{figure}[htbp]
\begin{minipage}{8.5cm}
 \includegraphics[height=7.4cm, width=8.5cm]{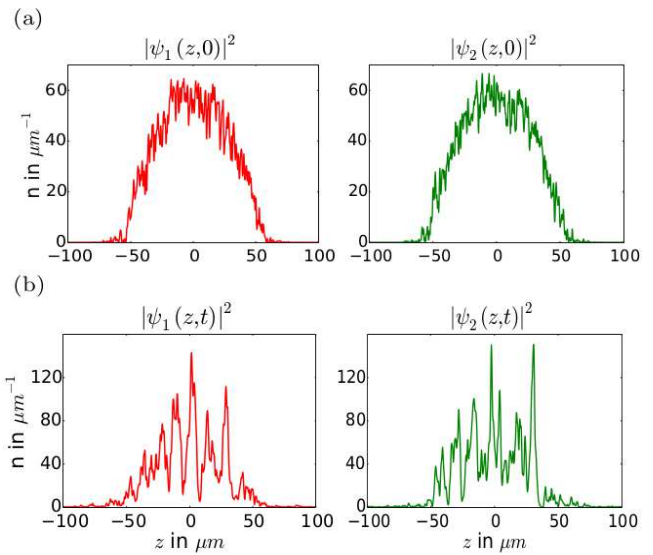}
\end{minipage}
\caption{\label{fig:wellenfkt} a) The densities of two stochastically independent solutions of equation \eqref{SMFE} $|\psi_1(z,0)|^2$ and $|\psi_2(z,0)|^2$ are plotted before ballistic expansion.
b) After free propagation for $t=\SI{22}{ms}$ these two densities $|\psi_1(z,t)|^2$ and $|\psi_2(z,t)|^2$  show the emergence of density ripples.
The values are $^{87}$Rb, $N=4400$, $T=\SI{31}{nK}$ and frequencies $\omega_\perp=2 \pi \cdot \SI{ 3000}{Hz}$, $\omega=2 \pi \cdot \SI{ 12}{Hz}$.}
\end{figure}

In the second step we model the ballistic expansion during the time period from switching off the traps until taking the absorption image.
Therefore we perform time propagation for $t=\SI{22}{ms}$.
After releasing from the traps the Bose gases inflate almost instantaneously in the transverse directions of initial strong confinement and are immediately diluted.
Thus we neglect self interaction during ballistic expansion and use the free Schrödinger equation as an approximation.
The time evolution of the c-fields $\Psi_1(x,y,z,t)$ and $\Psi_2(x,y,z,t)$ separates again into propagation in respective dimensions.
The free propagation of the Gaussian wave functions
\begin{align*}
\phi_{1,2} (x,y,t)= \left(\frac{m}{\pi \hbar \omega_\perp t^2}\right)^{1/2} \ee^{-\frac{m (1-i \omega_\perp t) }{2 \hbar \omega_\perp t^2} \left(\left(y \pm \frac{d}{2}\right)^2+x^2\right)}
\end{align*}
for the transverse directions are analytically well known.
In longitudinal direction we numerically apply $\psi_{1/2}(z,t)=\operatorname{exp}(-\frac{\mathrm{i}}{\hbar}\frac{\hat{p}^2}{2m} t)\psi_{1/2}(z,0)$.
Here we observe the formation of density ripples described in \cite{manz2010, imambekov}.
The spatial noise of the stochastic solutions $\psi_{1/2}(z,0)$ causes oscillations in the time developed densities $|\psi_{1/2}(z,t)|^2$.
Their mean wavelength and mean amplitude grow with increasing time.
In Fig. \ref{fig:wellenfkt} we show two densities in longitudinal direction before (a) and after (b) ballistic expansion.
In contrast to the transverse directions there is no noticeable spatial expansion of density along the longitudinal direction because of their loose initial longitudinal confinement.
The mean wavelength for the density oscillations (ripples) \cite{imambekov} is $\sqrt{\frac{2 \pi \hbar t}{m}}\approx \SI{10}{\micro m}$ with the above values which is consistent with the approximate distance of density maxima in Fig. \ref{fig:wellenfkt} (b).
\subsection{Absorption image}
In the last step we determine the interference pattern
\begin{align*}
 I(y,z) & =|\Psi_1 (y,z,t)+\Psi_2 (y,z,t)|^2 \\
   & =|\psi_1(z,t) \phi_1(y,t)+\psi_2 (z,t)  \phi_2(y,t)|^2
\end{align*}
by calculating the density of the coherent sum.
In experiments there is finite resolution of the optical devices, e.g. one pixel of the CCD camera is $\SI{3}{\micro m} \times \SI{3}{\micro m}$.
Therefore $I(y,z)$ has to be convolved with a point spread function
\begin{align}
\label{psf}
 \frac{1}{2 \pi \sigma^2} \ee^{-\frac{y^2+z^2}{2 \sigma^2}}
\end{align}
with $\sigma=\SI{3.4}{\micro m}$.
\begin{figure*}[htbp]
 \begin{minipage}{17.0cm}
   \includegraphics[height=5.5cm, width=17.0cm]{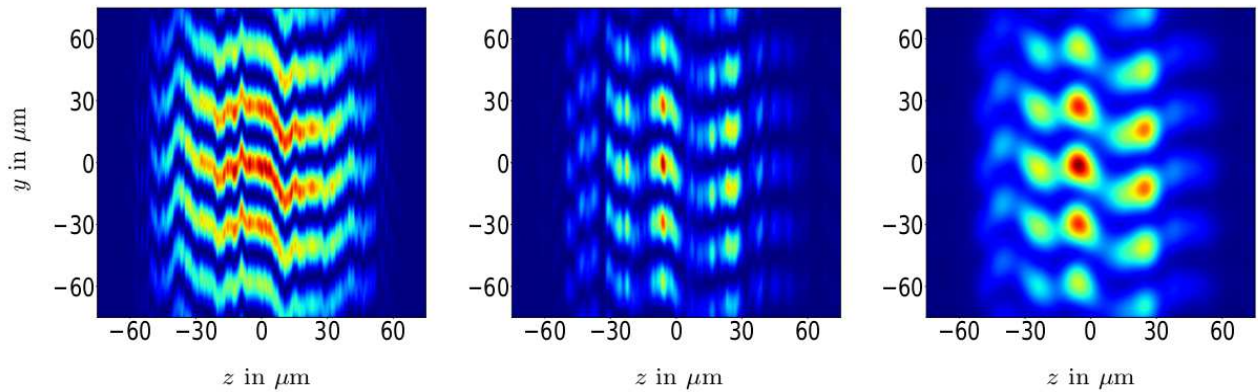}
 \end{minipage}
\caption{\label{fig:interferenzbild2} Left: One mathematical realization $I(y,z)$ without time propagation of the one-dimensional fields $\psi_{1/2}$, where the z-dependent relative phase shifts are pointed out.
Middle: The same realization with time propagation of the fields $\psi_{1/2}$ show the density ripple effect. Such a picture would be seen in experiments with infinite resolution of the optical devices.
Right: The same realization with time propagation of the fields $\psi_{1/2}$ and convolution with the point spread function from the experiment.
The values are $^{87}$Rb, $N=4400$, $T=\SI{31}{nK}$, $\omega_\perp=2 \pi \cdot \SI{ 3000}{Hz}$, $\omega=2 \pi \cdot \SI{ 12}{Hz}$, $d=\SI{3.5}{\micro m}$.}
\end{figure*}

\begin{figure*}[htbp]
 \begin{minipage}{17.0cm}
   \includegraphics[height=5.5cm, width=17.0cm]{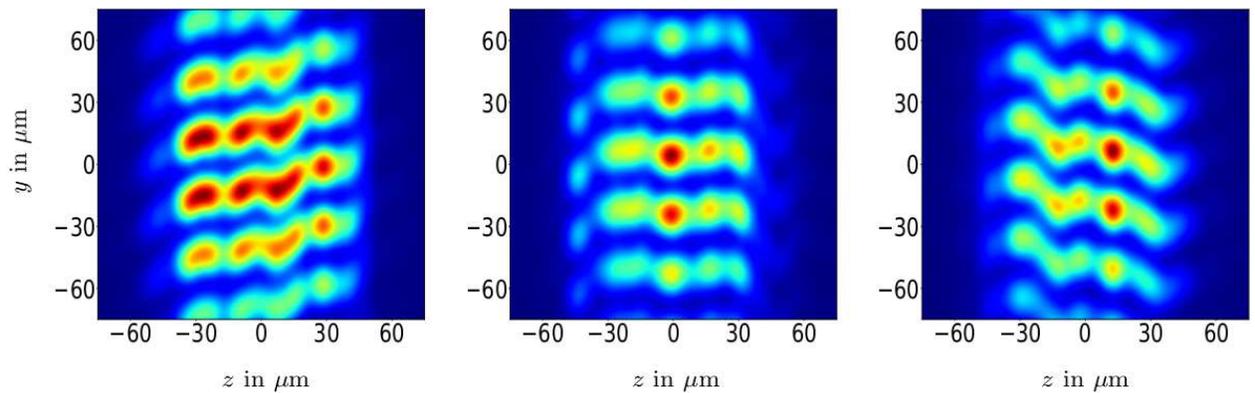} 
 \end{minipage}
\caption{\label{fig:interferenzbild3} Three more realizations of interference patterns, like Fig. \ref{fig:interferenzbild2}, right. }
\end{figure*}

In Fig. \ref{fig:interferenzbild2} we show the formation of one realization of an interference pattern calculated with the stochastic matter field equation.
We reveal the density-ripple effect during time propagation and the effect of the convolution with the point spread function \eqref{psf}.
Spatial details with size $\approx \SI{3}{\micro m}$ are coarse-grained which affects the density oscillations in $z$-direction.
The convolution has almost no effect in $y$-direction because spatial structures in this dimension are much larger than the width of the point spread function.
Three more examples of final interference patterns are shown in Fig. \ref{fig:interferenzbild3}.

If we compare our modeled interference patterns with the experimental ones \cite{schmied2008}, we find good agreement in all crucial features.
Characteristic properties like the distance between the stripes, the stochastic displacements of these stripes and intensity oscillations inside single stripes are realized in experimental and theoretical interference patterns.
The oscillatory structure inside single bright interference stripes can now be understood with density ripples which are washed out by finite resolution.

\section{Statistics of Contrast}
\label{sec:statis}
\begin{figure}[htbp]
\begin{minipage}{8.5cm}
\includegraphics[height=5.5cm, width=8.5cm]{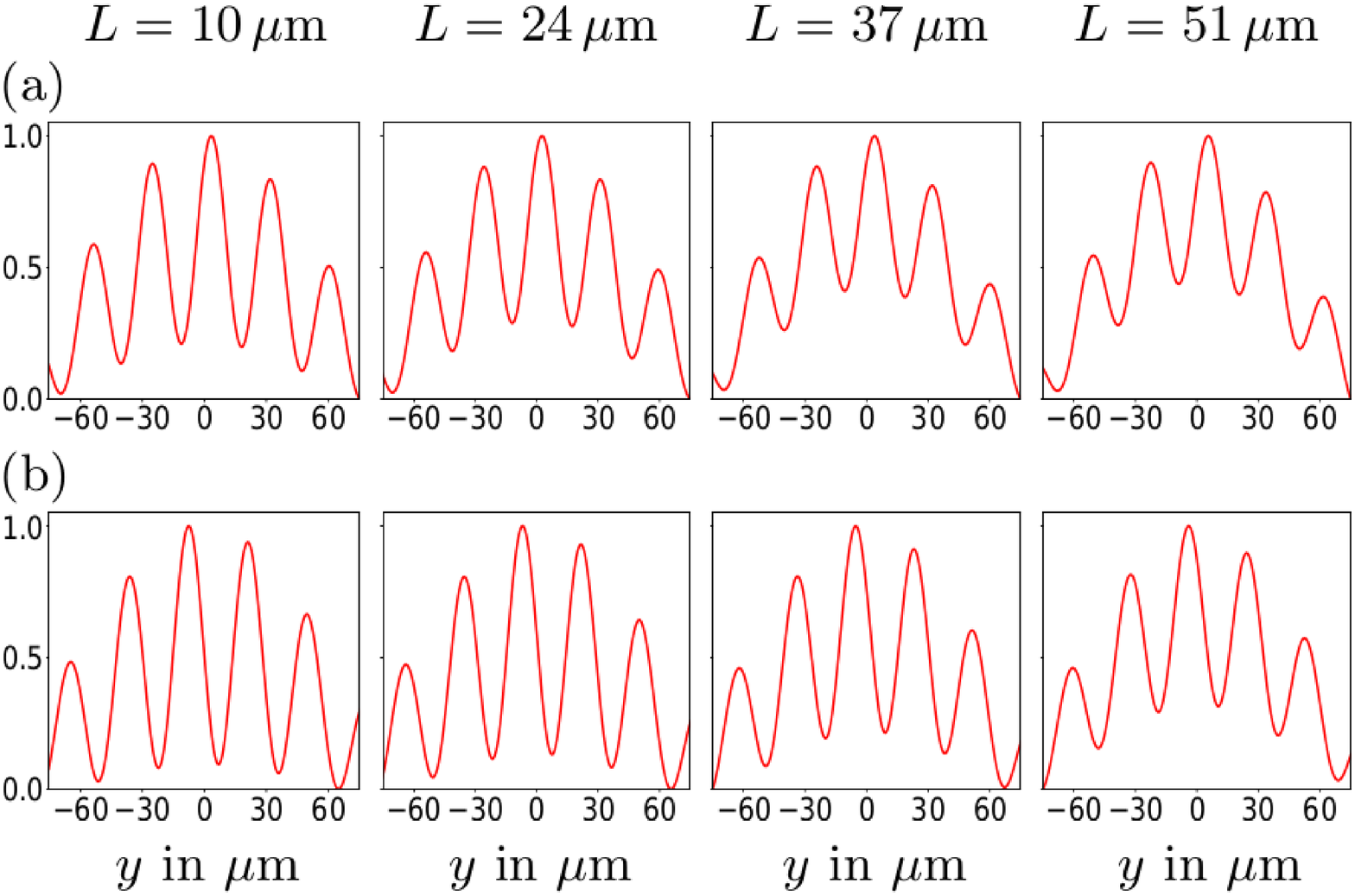} 
\end{minipage}
\caption{Normalized integrated densities from our theoretical interference patterns.
The upper row (a) shows integrated density functions $d(y)$ from the right pattern in Fig. \ref{fig:interferenzbild2} for respective integration intervals.
The lower row (b) presents another example.
For better graphic comparability, the functions are scaled between 0 and 1.
Without scaling, the contrast decreases with increasing integration length $L$.
The values are $^{87}$Rb, $N=4400$, $T=\SI{31}{nK}$, $\omega_\perp=2 \pi \cdot \SI{ 3000}{Hz}$, $\omega=2 \pi \cdot \SI{ 12}{Hz}$, $d=\SI{3.5}{\micro m}$. }
\label{fig:normdens}
\end{figure}

One single interference pattern is only one sample which does not permit access to probability distributions so in this section we examine statistical properties to evaluate the quality of our stochastic matter field equation and compare statistical quantities with experimental results.
The statistics of the solutions of the matter field equation determines the statistics of the integrated interference contrast, which is measurable in experiments.
So we can probe the properties of interacting many-particle quantum fields which are provided by our matter field equation.
In experiments \cite{schmied2007, schmied2008} the interference patterns were integrated symmetrically along the z-axis over various length scales $L=\SI{10}{\micro m}, \SI{24}{\micro m}, \SI{37}{\micro m}, \SI{51}{\micro m}$ respectively, and normalized.
In Fig. \ref{fig:normdens} we show integrated densities from our calculated interference patterns.

\begin{figure}[htbp]
\hspace{-14pt}
\begin{minipage}{8.5cm}
 \includegraphics[height=7.5cm, width=8.7cm]{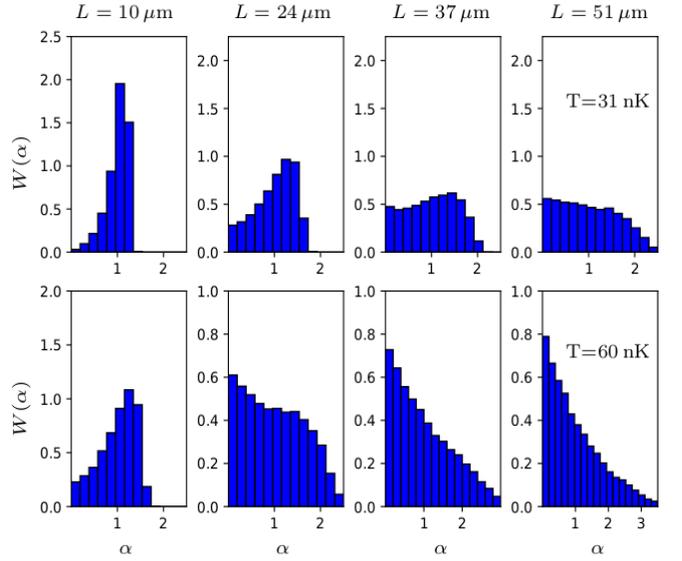} \\
\end{minipage}
 \caption{Histograms of the reduced contrasts $\alpha$ without measurement scheme ('in-situ', equation \eqref{visibility_insitu}) for almost the same values as in \cite{schmied2008}.
 Each histogram is made from $20000$ pairs of solutions of our stochastic matter field equation. The values are $^{87}$Rb, $N=4400$, $\omega_\perp=2 \pi \cdot \SI{ 3000}{Hz}$, $\omega=2 \pi \cdot \SI{ 12}{Hz}$, $d=\SI{3.5}{\micro m}$, upper row temperature $T=\SI{31}{nK}$ and lower row $T=\SI{60}{nK}$.}
 \label{fig:histschmied}
\end{figure}

\begin{figure}[htbp]
 \hspace{-22pt}
\begin{minipage}{8.5cm}
\includegraphics[height=7.5cm, width=9.0cm]{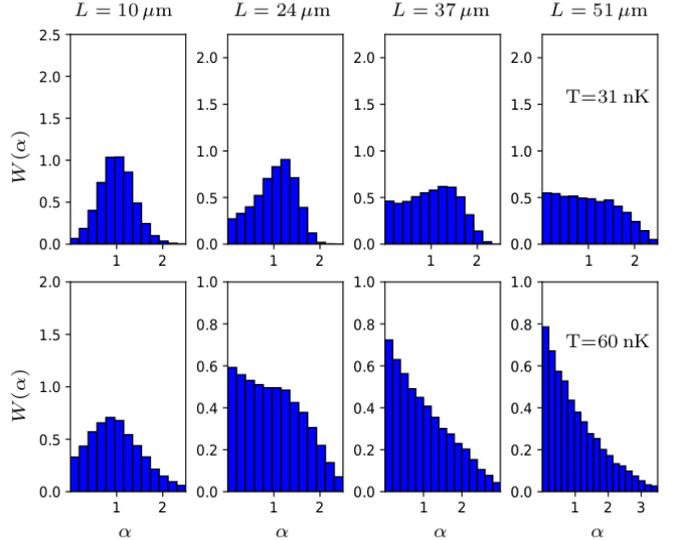}
\end{minipage}
 \caption{Histograms of the reduced contrasts $\alpha$ as in Fig. \ref{fig:histschmied}. Here the measurement process, that means ballistic expansion for $t=\SI{22}{ms}$ during free fall and convolution is included (equation \eqref{visibility}).}
 \label{fig:convschmied}
\end{figure}

Using the symbol ${\cal{F}}(\ldots)$ for the convolution with the point spread function \eqref{psf} yields the expression
 \begin{align}
 \label{intdensities}
  & d(y)  = \int_{-L/2}^{L/2} \limits \hspace{-0.1cm} dz \ \mathcal{F} \bigg( |\Psi_1 (y,z,t)+\Psi_2 (y,z,t)|^2 \bigg) \\
  \nonumber
     & \approx 2 \sqrt{\frac{m}{\pi \hbar \omega_\perp t^2}} \ee^{-\frac{ m y^2}{\hbar \omega_\perp t^2}} \bigg(C(L)+|A(L)| \cos \left(Q y+ \chi(L) \right) \bigg) 
  \end{align}
of the integrated density function. The wave vector $Q=\frac{m d}{h t}$ determines the distance between the interference stripes which is, in our case, $\frac{2 \pi}{Q} \approx 30 \operatorname{\mu m}$.
We assert that the integrated densities have a Gaussian envelope with a large standard deviation compared to the size of the interference patterns because $\sqrt{\frac{\hbar \omega_\perp t^2}{2 m}} \approx \SI {120}{\micro m}$.
The parameter
\begin{align*}
  & C(L)  = \frac{1}{2} \int_{-L/2}^{L/2} \limits \hspace{-0.1cm} dz \ \mathcal{F} \bigg( |\psi_1 (z,t)|^2 + |\psi_2 (z,t)|^2 \bigg) \\
 \end{align*}
is a constant random variable with respect to $y$.
The quantities $|A|$ and $\chi$ are absolute value and phase of a complex random variable
 \begin{align}
 \label{visibility}
  & A(L)  = |A| \ee^{i \chi} =  \int_{-L/2}^{L/2} \limits \hspace{-0.1cm} dz \ \mathcal{F} \bigg( \psi_1^* (z,t) \psi_2 (z,t) \bigg) \text{,}
 \end{align}
which is of central importance.
The amplitude $|A|$ has the meaning of visibility  and $\chi$ defines the phase of the cosine function.
For fixed $L$, amplitude and phase are random variables and differ in every realization of the experiment and, accordingly, for every pair of solutions of our stochastic matter field equation.
In the theoretical description we do not need to evaluate interference patterns but can directly use equation \eqref{visibility}.
For stochastically independent solutions (independent Bose gases) the phase of $A$ is uniformly distributed, the mean value vanishes $\langle A \rangle=0$ and so statistics with $|A|^2$ is feasible.
We use the normalized interference contrast \cite{schmied2008}
\begin{align*}
 \alpha(L)=\frac{|A(L)|^2}{\langle |A(L)|^2 \rangle} \text{.}
\end{align*}
In many repetitions we calculate independent pairs of solutions of the matter field equation, calculate a set of events $\{ \alpha(L) \}$ and represent the results in a histogram for various fixed temperatures and integration lengths. 
In Fig. \ref{fig:histschmied} and Fig. \ref{fig:convschmied} these distributions of the normalized interference contrast calculated with independent solutions of the matter field equation are shown.

First we omit the measurement process and calculate the visibility
\begin{align}
\label{visibility_insitu}
 A(L)=\int_{-L/2}^{L/2} \limits \hspace{-0.1cm} dz \ \psi^*_1(z) \psi_2(z)
\end{align}
without ballistic expansion $\psi^{1/2}(z)=\psi^{1/2}(z,0)$ and without convolution $\mathcal{F}$.
In Fig. \ref{fig:histschmied} we show the corresponding results for the normalized contrasts (see also \cite{sigi2013}) and see very good agreement with Luttinger liquid theory and experiments \cite{schmied2008}.

Second, in Fig. \ref{fig:convschmied} we respect the measurement process, that means ballistic expansion during free fall and convolution due to finite resolution of the optical devices is included (equation \eqref{visibility}).
The differences between Fig. \ref{fig:histschmied} and Fig. \ref{fig:convschmied} are caused by the measurement scheme which effects mainly the `short' sample $L= \SI{10}{\micro m}$.
Numerical calculations show that the deviations are mainly caused by the free time propagation during ballistic expansion. 



\section{conclusion}
In this paper we have presented a c-field method for the canonical ensemble to describe an interacting Bose gas at finite temperature with fixed particle number.
This stochastic matter field equation is rigorously exact for the ideal gas, and we include self interaction in mean-field approximation.
The crucial point is that we use colored noise which facilitates a holistic treatment of the Bose gas with one single stochastic field.
In contrast to other methods, cutoff energies or separated treatment of condensate and non-consensed particles is omitted in a natural way.
As an example, we applied this equation to interference experiments with one-dimensional interacting Bose gases and see very good agreement with crucial experimental results.
We emphasize that there is no dimensional restriction in the derivation of our stochastic matter field equation, so it is valid also in three dimensions.
Moreover, it enables efficient numerical implementation in arbitrary dimensions.
Expectation values of products of stochastic fields exhibit correlation functions of arbitrary order and therefore in principle our equation enables access to the full quantum state.
Due to technological and experimental developments tests of the presented theory on higher order correlation functions (measured in \cite{langen207}) are possible.

In this paper, we discussed equilibrium physics, but the method is also suitable for the huge subject of non-equilibrium physics.
To calculate equilibrium states, the specific value of the damping parameter $\Lambda$ is irrelevant for $\Lambda t \gg 1$.
Then, with arbitrary $\Lambda > 0$ our equation will force the matter field into a stationary state: on average the properties of this fluctuating field are constant and the distribution functional is constant.

Further investigation should focus on non-equilibrium physics and how to adjust the damping parameter $\Lambda$ to systems of interest. 
In recent years fascinating research on relaxation dynamics or thermalization of many-body systems \cite{nowak2014, zill2015, gogolin2016, langen2016} appeared and it could be an interesting application for the presented method.
For example, depending on the choice of $\Lambda$, it would be interesting to see whether our theory also describes pre-thermalization and non-thermal fixed points.
Further application could contribute to current issues of dynamics and dissipation of solitons in Bose gases \cite{cockburn2011_2, gallucci2016}.
Polariton condensates could also be an interesting field of research.
Moreover, application to vortex dynamics and dissipation of vortices appears worthwile \cite{yan2014, berloff2014, white2014}.
In that case an additional stirring potential could be included to the stochastic matter field equation to simulate the stirring lasers which create the vortices.


\begin{acknowledgments}
We are grateful to Sigmund Heller for his earlier contributions.
\end{acknowledgments}

\appendix
\section{Lindblad equation and Fokker-Planck equation for the grand canonical ensemble}
\label{sec:anhang0}
Time propagation of the density operators for an ideal gas are phenomenologically governed by Lindblad master equations
\begin{align*}
 \frac{\dif \hat{\varrho}_i}{\dif t} = -\frac{\mathrm{i}}{\hbar} \left[ \hat{H}_i, \hat{\varrho}_i \right]  &+ \frac{\gamma_i}{2} n_i \left( \left[ \hat{a}_i^\dagger \hat{\varrho}_i, \hat{a}_i  \right] + \left[ \hat{a}_i^\dagger , \hat{\varrho}_i \hat{a}_i  \right] \right) \\
  &+\frac{\gamma_i}{2} (n_i+1) \left( \left[\hat{a}_i \hat{\varrho}_i, \hat{a}_i^\dagger \right] + \left[ \hat{a}_i, \hat{\varrho}_i \hat{a}_i^\dagger \right] \right)
 \end{align*}
for each mode $i$ \cite{Breuer2007,carmichael1998}.
At equilibrium the mean quantum numbers are $n_i=\frac{1}{e^{\beta (E_i-\mu)}-1}$.
Plugging in equation \eqref{definitio} into the Lindblad equation yields
\begin{align}
\label{fpegc}
\begin{split}
& \frac{\partial P_i(z_i, z_i^*)}{\partial t} = \left( \left(\frac{\gamma_i}{2}+\mathrm{i} \frac{(E_i-\mu)}{\hbar} \right) \frac{\partial }{\partial z_i} z_i  \right. \\
 & \ \ \  \left.  +\left(\frac{\gamma_i}{2}-\mathrm{i} \frac{(E_i-\mu)}{\hbar} \right) \frac{\partial }{\partial z_i^*} z_i^* +\gamma_i n_i  \frac{\partial^2 }{\partial z_i^* \partial z_i} \right) P_i(z_i, z_i^*) \ \text{,}
\end{split}
 \end{align}
which are the Fokker-Planck equations for the P-functions $P_i$.

\section{Calculation of the canonical partition function}
\label{sec:anhangdetails}

We calculate the canonical partition function
\begin{align}
\begin{split}
 Z_N & = \operatorname{Tr} \left(\ee^{-\beta \sum_i \limits E_i \hat{a}^\dagger_i \hat{a}_i} \hat{\Pi}_N \right) \\
  &= \int \dif ^2 \{ z \} \ \prod_k \tilde{P}_k(z_k,z_k^*) \langle z_k | \hat{\Pi}_N |z_k \rangle 
\end{split}
\end{align}
using the P-functions of the unnormalized exponential operators, equations \eqref{pfunctionunnorm} and \eqref{Pfunction} with $E_i-\mu \rightarrow E_i$.
The expectation value of the projection operator with respect to coherent states
\begin{align}
\label{epo}
 \prod_k \limits \langle z_k | \hat{\Pi}_N |z_k \rangle = \ee^{-\sum_k \limits |z_k|^2} \frac{\left( \sum_k \limits |z_k|^2 \right) ^N}{N!}
\end{align}
is well known \cite{mollow1968}.
The combination of equations \eqref{pfunctionunnorm} and \eqref{epo} gives the canonical partition function.

The correlation function of first order can be written as
\begin{align*}
\begin{split}
 &\langle \hat{a}^\dagger_j \hat{a}_j\rangle_N = \int \dif^2 \{ z \} \ \prod_k \limits \langle z_k |   \hat{a}^\dagger_j  \hat{a}_j \hat{\Pi}_N |z_k \rangle \frac{\tilde{P}_k(z_k,z_k^*)}{Z_N} \\
 \end{split}
\end{align*}
With equations \eqref{pfunctionunnorm}, \eqref{Pfunction} and \eqref{epo} we end up with
\begin{align*}
\begin{split}
 \langle \hat{a}^\dagger_j \hat{a}_j\rangle_N &= \frac{ 1}{Z_N} \prod_j \limits \ee^{\beta E_j} \cdot \\
 &\ \ \ \ \ \ \   \int \dif^2 \{ z \} \frac{|z_j|^2 \ee^{-\sum_l \limits |z_l|^2 \ee^{\beta E_l}}}{(N-1)!}\left( \sum_l \limits |z_l|^2 \right)^{N-1}  \\
 &=:\frac{1}{Z_N}  \prod_j \ee^{\beta E_j} \int \dif^2 \{ z \} |z_j|^2 \ W_{N-1} (\{ z \} )
\end{split}
\end{align*}
in which we rediscover the weight function with index $N-1$.
In short notation this correlation function is
\begin{align}
\label{firstorder2}
 \langle \hat{a}^\dagger_j \hat{a}_j\rangle_N =\frac{ \int \dif^2 \{ z \} \ |z_j|^2 \ W_{N-1} (\{ z \} )}{\int \dif^2 \{ z \} \ W_N(\{ z \})} \ \text{.}
\end{align}
Similarly, the second order correlation functions can be obtained straightforwardly
\begin{align}
\label{secondorder2}
 \langle \hat{a}^\dagger_j \hat{a}^\dagger_k  \hat{a}_l  \hat{a}_m \rangle_N = \frac{ \int \dif^2 \{ z \} \ z_j^* z_k^* z_l z_m W_{N-2} (\{ z \} )}{\int \dif^2 \{ z \} \ W_N(\{ z \})}
\end{align}
and in the same way we can get expectation values for normal-ordered products of creation and annihilation operators of arbitrary order
\begin{align}
\label{generalorder2}
 \langle \underbrace{ \hat{a}^\dagger_j \hat{a}^\dagger_k \ldots \hat{a}_l  \hat{a}_m }_{\substack{2M}}\rangle_N &= \frac{ \int \dif^2 \{ z \} \ z_j^* z_k^* \ldots z_l z_m \ W_{N-M} (\{ z \} )}{\int \dif^2 \{ z \} \ W_N(\{ z \})} 
\end{align}
with $N\geq M$.
The denominators in equations \eqref{firstorder2}, \eqref{secondorder2} and \eqref{generalorder2} arise from the partition function $Z_N$ and lead to normalization.
We mention the important recurrence property of the weight functions
\begin{align*}
 W_N=\frac{1}{N} \left( \sum_k \limits |z_k|^2 \right) W_{N-1} \ \text{,}
\end{align*}
therefore it is sufficient to know only $W_{N}(\{ z \})$, and, for instance, equation \eqref{firstorder2} can be written as equation \eqref{firstorder3}.

\section{Densities in position representation}
\label{sec:anhangdensities}
The densities in position space are the correlation functions of first order.
For the grand canonical ensemble, equation \eqref{firstorder} becomes
\begin{align*}
n(x) &= \langle \hat{\Psi}^\dagger(x) \hat{\Psi}(x) \rangle_\infty= \frac{\displaystyle \int {\cal{D}} [\psi,\psi^*] \  |\psi(x)|^2 P[\psi,\psi^*]}{\displaystyle \int {\cal{D}} [\psi,\psi^*] \  P[\psi,\psi^*]} \\
 & = \langle \hspace{-0.2cm} \langle \ |\psi(x)|^2 \ \rangle \hspace{-0.2cm} \rangle_\infty\ \text{,}
\end{align*}
where the denominator ensures normalization because we used an unnormalized P-functional.
With sample functions governed by the distribution $P[\psi,\psi^*]$ this expectation value is an average over many realizations $n(x)= \langle \hspace{-0.2cm} \langle \ |\psi(x)|^2 \ \rangle \hspace{-0.2cm} \rangle_\infty$ . 
Very similar, equations \eqref{firstorder2}, \eqref{firstorder3} and \eqref{firstorder3a} for the canonical ensemble give
\begin{align*}
 n(x) &= \langle \hat{\Psi}^\dagger(x) \hat{\Psi}(x) \rangle_N= \frac{\displaystyle \int  {\cal{D}} [\psi,\psi^*] \  |\psi(x)|^2 W_{N-1}[\psi,\psi^*]}{\displaystyle \int  {\cal{D}} [\psi,\psi^*] \  W_N [\psi,\psi^*]} \\
 & = N \Bigg\langle \hspace{-0.2cm} \Bigg\langle  \frac{|\psi(x)|^2}{\langle \psi | \psi \rangle}  \Bigg\rangle \hspace{-0.2cm} \Bigg\rangle_N 
\end{align*}
in position representation.

\section{Grand canonical ensemble}
\label{sec:anhang1}
Here we give some intermediate steps for the calculation that the P-functional \eqref{Pfunctional} is a solution of the stationary Fokker-Planck equation \eqref{fpefunctional}.
The functional derivative of the drift parameter is
\begin{align*}
 \frac{\delta (A_\infty \psi(x))}{\delta \psi(x)}=\frac{\mathrm{i}}{\hbar} \frac{\delta ( (H-\mu) \psi(x))}{\delta \psi(x)}+\frac{1}{2} \big\langle x \big| \gamma \big| x \big\rangle \ \text{.}
\end{align*}
The functional derivatives of the P-functional are
\begin{align*}
 \frac{\delta P}{\delta \psi(x)} &= -P \big\langle \psi \big| \ee^{\beta (H-\mu)}-1 \big| x \big\rangle \\
 \frac{\delta^2 P}{\delta \psi(x) \delta \psi^*(x')} &= P   \big\langle x' \big| \ee^{\beta (H-\mu)}-1 \big| \psi \big\rangle \big\langle \psi \big| \ee^{\beta (H-\mu)}-1 \big| x \big\rangle \\
 & \ \ \ \ \ \ \ \ \ -P \big\langle x' \big| \ee^{\beta (H-\mu)}-1 \big| x \big\rangle  \ \text{.}
\end{align*}
The diffusion term of the Fokker-Planck equation is
\begin{align*}
 \int \dif x & \int \dif x' \  B_\infty(x,x')  \frac{\delta^2 P}{\delta \psi(x) \delta \psi^*(x')} \\
 &=P \big\langle \psi \big| \gamma (\ee^{\beta (H-\mu)}-1) \big| \psi \big\rangle - P  \int \mbox{d} x \big\langle x \big| \gamma \big| x \big\rangle \ \text{.}
\end{align*}
\ \\

\section{Canonical ensemble}
\label{sec:anhang2}
With functional derivatives for the drift
\begin{align*}
 \frac{\delta \big(A_N \psi(x)\big)}{\delta \psi(x)} &=\big\langle x \big|  \frac{\mathrm{i}}{\hbar}  H \big| x \big\rangle-\big\langle x \big| \frac{\Lambda}{2} \frac{N}{\big\langle \psi \big| \psi \big\rangle} \ee^{-\beta H} \big| x \big\rangle \\
 &\ \ + \frac{N}{2}  \frac{ \big\langle \psi \big| x \big\rangle }{\big( \big\langle \psi \big| \psi \big\rangle \big)^2}  \big\langle x \big| \Lambda \ee^{-\beta H} \big| \psi \big\rangle+\big\langle x \big|\frac{\Lambda}{2} \big| x \big\rangle 
\end{align*}
and the probability density functionals
\begin{align*}
 &\frac{\delta W_N[\psi,\psi^*]}{\delta \psi(x)} =W_N \left( \frac{N}{\big\langle \psi \big| \psi \big\rangle} \big\langle \psi \big| x \big \rangle -\big\langle \psi \big| \ee^{\beta H} \big| x \big \rangle  \right) \\
\end{align*}
\begin{align*}
 &\frac{\delta^2 W_N[\psi,\psi^*]}{\delta \psi(x) \delta \psi^*(x')} = W_N \Bigg( \frac{N^2-N}{\big( \big\langle \psi \big| \psi \big\rangle\big)^2} \big\langle x' \big| \psi \big\rangle \big\langle \psi \big| x \big\rangle  \\
 &  - \frac{N}{\big\langle \psi \big| \psi \big\rangle } \big\langle x' \big| \ee^{\beta H} \big| \psi \big \rangle \big\langle \psi \big| x \big\rangle +\frac{N}{\big\langle \psi \big| \psi \big\rangle} \delta(x-x')  \\
 &  - \frac{N}{\big\langle \psi \big| \psi \big\rangle } \big\langle x' \big| \psi \big \rangle \big\langle \psi \big| \ee^{\beta H}  \big| x \big\rangle + \big\langle x' \big| \ee^{\beta H}\big| \psi \big\rangle \big\langle \psi \big| \ee^{\beta H}\big| x \big\rangle  \\  
 &  - \big\langle x' \big| \ee^{\beta H} \big| x \big\rangle \Bigg)
\end{align*}
one can prove that $W_N$ is a solution of the canonical Fokker-Planck equation \eqref{fpefunctional} with choices \eqref{canparameter}.


\bibliography{literature_arxiv.bib}

\end{document}